**Title: Peer to peer learning platform optimized with machine learning**

Vikram Anantha
Lexington High School

**Abstract**

HELM Learning (Helping Everyone Learn More) is the first online peer-to-peer learning platform which allows students (typically middle-to-high school students) to teach classes and students (typically elementary-to-middle school students) to learn from classes for free. This method of class structure (peer-to-peer learning) has been proven effective for learning, as it promotes teamwork and collaboration, and enables active learning. HELM is a unique platform as it provides an easy process for students to create, teach and learn topics in a structured, peer-to-peer environment. Since HELM was created in April 2020, it has gotten over 4000 student sign ups and 80 teachers, in 4 continents around the world. HELM has grown from a simple website-and-Google-Form platform to having a backend system coded with Python, SQL, JavaScript and HTML, hosted on an AWS service. This not only makes it easier for students to sign up (as the students' information is saved in an SQL database, meaning they can sign up for classes without having to put in their information again, as well as getting automated emails about their classes), but also makes it easier for teachers to teach (as supplemental processes such as creating Zoom links, class recording folders, sending emails to students, etc. are done automatically). In addition, HELM has a recommendation machine learning algorithm which suggests classes and subjects students would enjoy taking, based on the previous classes a student has taken. This has created an easier experience for students to sign up for classes they are interested in.

## 1. Introduction

Peer-to-peer learning is a process by which individuals exchange knowledge and experience with each other, and specifically, when students learn from students in a formal or informal way. According to Boud and colleagues (Boud et al., 2014; Boud et al., 2001), peer learning involves sharing of ideas and concepts between individuals which can be mutually beneficial. Thus, the students, when asked to teach materials, can become proficient in a topic by planning learning activities and expressing concepts to listeners and collaborating in the discussions (Boud et al., 2001). There are many benefits of peer-to-peer learning exercises which include fostering teamwork and collaboration and in this way social skills, critical enquiry and reasoning skills, clear expression of knowledge and skills, and managing and enabling active learning (Boud et al., 2014). Further, in a peer-to-peer learning environment, there is mutual trust between the teacher and students, especially because the teachers are also students. Further, students tend to learn and remember more information when they interact with their peers rather than being taught by a teacher. There are many examples of peer-to-peer education websites that allow anyone to create “content” for their own consumption (e.g., Quizlet allows middle and high school students to create flashcards or quizzes to study for exams) or for other select group of people (e.g., Tynker, a platform that allows students to self-learn coding). These are however, very specific in the educational content covered or in the way that which students learn material (e.g., practice quizzes, study for exams, but no synchronous lectures).

There are online education platforms such as Khan Academy or AOPS, but these are structured platforms, where course materials are taught in terms of lectures by adults, and students can only progress from one lecture to another when age or grade appropriate or deemed ready by an instructor. These are not peer-to-peer learning platforms. Also, relevant to this project, these platforms do not have class recommendation algorithms that would enable kids to get exposed to new topics or subjects that are not part of a structured curriculum.

One of the potential disadvantages of peer learning is that over time, the informal arrangements of learning together may disintegrate as people have better/more interesting things to do with their time. However, during the COVID-19 pandemic, all outside of school activities came to a halt and students were forced to retreat to their homes to continue learning, a huge non-health impact of the pandemic (Guppy et al., 2021). The present project describes the development and deployment of an free online K-12 peer-to-peer learning platform HELM (Helping Everyone Learn More), which enables any student to teach other students subjects or topics that they have mastered. As of September 2022, the platform hosts 80 classes with over 4,100 students, making it a very valuable resource for outside school education.

As HELM has grown in popularity among students and peer “student teachers”, the platform has evolved from a simple website to a sophisticated and automated peer-to-peer software platform that is enabled by machine learning. The HELM platform is built using a scalable architecture using a LAMP (Linux, Apache, MySql, Python) software stack, and hosted in the “cloud” on AWS (Amazon Web Services), so it is available to everyone online. The architecture of the platform also enables detailed data collection about the classes, students’ and teachers’ interactions to be stored in a manner that can be used by sophisticated machine learning systems. The HELM platform’s Machine Learning (ML) system currently uses this data to make class recommendations to students on completely new topics by predicting what they might have an interest in.

The HELM platform is unique because it makes the process really easy for K-12 kids to create, teach and learn topics in a structured, peer-to-peer and collaborative environment. With HELM, students can learn from one another on a variety of topics, and its machine learning system enables kids to be exposed to new topics by predicting their interests. It is the first peer-to-peer learning platform of its kind, and can grow to become an invaluable after-school learning resource for kids around the world. This would be especially important in breaking down the geographical barriers of access to quality after-school programs in minority and low-income communities around the US and the world.

## 2. What is the HELM peer-to-peer learning platform

HELM (Helping Everyone Learn More) is the first online peer-to-peer learning platform for K-12 kids, which enables any student to connect with and teach other students subjects or topics that they have mastered. Figure 1 shows the main website (https://helmlearning.com/) that is provided by the HELM platform. With the HELM platform, (typically) high school students teach their peers (other high school or younger kids) 60 - 90 minute classes on topics they have mastered. The typical class on HELM serves as a great introduction to a topic - a wide range of topics are available such as Calligraphy, Astronomy, Music theory, and Python programming. Figure 2 shows a sample of the classes currently available via the HELM platform today (https://helmlearning.com/our-classes.html). Students taking the classes get introduced to a new subject by their peers and advance their knowledge by learning from other students. The high school students teaching classes get better at the subject by teaching it to other kids and get Community Service hours for their work via HELM. Figure 3 shows some of the HELM teachers (typically high school students) who have taught classes through the platform (https://helmlearning.com/our-team.html).

## 3. Evolution of the HELM platform

The first version of the HELM platform was created during the pandemic as a simple static website which listed a few classes that students could “sign up” for via a Google Form. This process required a lot of manual steps such as setting up and coordination of the classes, and updating the website with new information as needed. Very quickly, as HELM grew in popularity among students and peer “student teachers”, the platform has evolved to be much an automated, maintainable and scalable peer-to-peer learning platform.

## 4. HELM Software platform design

The HELM platform is built using a scalable architecture using a LAMP (Linux, Apache, MySql, Python)(*LAMP (software bundle)*, 2022) software stack, and hosted in the “cloud” so it is available to everyone online. The architecture of the platform also enables detailed data collection about the classes, students' and teachers' interactions to be stored in a manner that can be used by sophisticated machine learning systems. The HELM platform's Machine Learning (ML) system currently uses this data to make class recommendations to students on completely new topics by predicting what they might have an interest in. Figure 4 shows the architecture of the HELM software platform, which is modular and scalable. The Software platform is separated into the "Frontend" system (which has the user interface), and the "Backend system" (which has all the intelligence) and the data layer (which stores all the data). The frontend and backend systems are cleanly separated and communicate via a RESTful API ("What is REST," 2022) using the standard JSON web services (*JSON-WSP*, 2021).

As shown in Figure 4, the "Backend system" is hosted in the cloud and has the HELM *software services* (code) layer and a separate *data layer* (HELM data store).

- All the HELM software services (backend) are hosted (run) on the Amazon Web Services' (AWS (https://aws.amazon.com/)) Elastic Cloud Compute service (EC2, (https://aws.amazon.com/ec2/)). All the components of the HELM software services are written in the Python programming language. A "small" EC2 server instance is currently used to host (run) the HELM software services.
- The HELM software services do not themselves store or have any data within them. Instead the HELM Data store is separated from the code and stored in the combination of components. The most important data layer component is the HELM database which has several data tables stored in a MySQL database (this is a relational database)(*MySQL*, 2022). An instance of the AWS Relational database service (RDS (https://aws.amazon.com/rds/) is used to host the HELM (MySQL) database to get scalability and redundancy.
- The HELM software services also use and store data files such as images and videos. For the data files such as images (such as the HELM logo, teacher photos etc.) that are used for internal purposes by the HELM software (for example: to generate the HELM webpages), we use AWS S3 (Simple storage service (https://aws.amazon.com/s3/)).
- For the data files that are shared externally with others (such as the class recordings), the HELM software uses Google Drive storage.

The users of HELM (such as teachers, students) interact with HELM's system through its "front end system". This includes the HELM website and emails that are dynamically generated by the backend system.

- The HELM system's website (each webpage) and all the HELM emails are dynamically generated by the "backend system" based on the classes currently available, and the interaction or behavior of the user.
- Each HELM webpage and each email is implemented in HTML / CSS / JavaScript code to make it dynamic.

## 5. HELM Software Services and HELM Databases

The HELM software services (backend) are hosted on a secure AWS EC2 server instance. Figure 5 shows all the HELM software services and its interaction with different types of HELM users (students, teachers and the HELM classes review panel). The primary HELM Software services include the Student signup service, Teacher signup service, Class setup and approval service, Class matching, coordination and logistics service, and the Data collection service. These software services form the core of the HELM peer-to-peer platform and enable teachers to add new classes, schedule them, and enables students to take the classes of their interest. Next, these software services are described in greater detail.

### 5.1. HELM Student signup service

The Student signup service provides the webpages and emails required for students to sign up for HELM classes. Figure 6 shows the algorithm and the data flow for the Student signup service when a new student registers with HELM. As shown in the figure, a student starts by entering basic information in the HELM student sign up webpage (front end), and this data is processed by the Student signup service, which determines if this is a new student by querying the Students table in the HELM database. For a new student, the Student signup service then dynamically generates the webpage that allows the student to fill the additional student demographic and profile information, which is then used to create a new student in the HELM database (Student table).

The Student Signup service additionally processes the information about the class this student has expressed an interest in, and adds the appropriate link between the student and the class in the Classes_to_students table of the HELM database. Finally, the Student signup service dynamically generates a Welcome email, including the personalized information about the student, the class they are registering for, and any additional instructions about the class. It sends this Welcome email to the student to complete this student's sign up process. The sign up process also handles various scenarios, including error cases that are not explicitly shown in Figure 6.

**5.2. HELM Teacher signup service**

The Teacher signup service gives teachers a process to register with HELM and provide the information required for their class to be approved on the platform. Figure 7 shows the algorithm and the data flow for the Teacher signup service when a new teacher wants to teach a new class via HELM. As shown in the figure, a teacher starts by entering basic information in the HELM Teacher sign up webpage (front end), and this data is processed by the Teacher signup service, which determines if this is a new teacher that wants to teach a new class - by querying the Teachers table and the Classes table in the HELM database. For a new class, the Teacher signup service then dynamically generates the webpage that allows the teacher to fill the additional teacher profile information (including a Google folder link to their photo), and class details (including the Class description, grade level etc.), which is then stored in the Cache table of the HELM database. The teacher photo provided by the teacher is stored in the HELM S3 storage for later use.

At this point, the new Class information is ready to be reviewed by the HELM Classes review panel (a panel of experts who ensure the quality control of the classes available through the platform). Appropriate emails are generated by the Teacher signup service and sent - an email to the HELM Classes review panel to review and approve this new class request, and an acknowledgement email is sent to the teacher that a review is in progress.

**5.3. HELM Class setup and approval service**

The "Class setup and approval" service automates the review and approval of a new class that has been uploaded by a teacher. Figure 8 shows an example of the approval process algorithm and data flow. As shown in the figure, the Class approver reviews the information provided to them via an email from HELM. They click a link in the email to approve the class, and provide "Class tags" that appropriately categorize this class. The tags corresponding to this class can help the HELM machine learning system in making personalized class recommendations (as described later in this paper).

After the class and teacher have been approved to be part of HELM's classes, the HELM service copies the relevant data from the HELM database's Cache table into the Teacher's table (if its a new teacher), and the Classes table. The appropriate links entries are also added into the Classes_to_teachers table to indicate that this teacher will be teaching this class. The "Class tags" are also added to the "Tags" table (if there are new tags), and appropriate links added to the Classes_to_tags table of the HELM database.

The HELM service also generates an email dynamically to the teacher to inform them that their class is now ready to be shown on the HELM website. Once the teacher indicates that they are ready to teach this class (by clicking on an auto-generated smart weblink that is in their email), the HELM service automatically creates a Zoom link (via the Zoom integration built into the HELM service). Correspondingly, the information for this class is updated in the Classes table in the HELM database indicating that this class should now be made available, and the Zoom link is available for use.

At this point, the HELM service (Class setup and approval service) generates the appropriate updates for the HELM webpages - this includes the teachers webpage that includes the new teacher's information (including their photo that was previously stored in Google Drive), and the classes webpage that includes information such as the class description, Zoom link, and the schedule of the class. This entire process and associated error scenarios are automatically handled by HELM's Class setup and approval software service, making it fairly easy for teachers to add and set up new classes that they would like to teach on the HELM platform. The manual step of having the HELM class review panel ensures that the appropriate quality of classes is maintained on the HELM peer-to-peer learning platform.

**5.4. HELM Class matching and class coordination service**

When a class is available for students to take (either a new class that has just been setup, or a new session for an existing class is made available), HELM's "Class matching and coordination" service implements a sequence of steps to ensure that the class runs smoothly. Figure 9 shows the example of the algorithm and data flow for a new class that has just been setup. As shown in the figure, this HELM service dynamically generates some marketing materials that can be used to promote this class. This includes an auto-generated QR code that encodes all the class info that can be easily shared and scanned to sign for this class; and an auto-generated marketing flier that promotes the class (and can be printed and / or shared easily). The sharable marketing material for the class is stored in a shareable Google Drive folder automatically, with a link to this material updated in the Classes table of the HELM database.

The HELM Class matching & coordination service then automatically generates & sends an email to the class-teacher informing them that their class is available on the HELM website, and they can promote it to their network with the marketing materials provided by HELM. This email is sent (with slight variations) automatically to the teacher a few times.

The HELM "Class matching & coordination" service then invokes the Machine Learning (ML) system to generate a "target list" of students for whom this class is a "recommended class" (in the top 3 recommended classes for the student). If the class has been previously taught, the ML system can rely on previous class and student history to make the prediction (as discussed in the ML section of this paper). The HELM service then dynamically generates an email that is sent to this "target list" of students informing them that a new class is available to them, and provides details of this class so they can sign up for it. This email is sent multiple times to the students to encourage students to sign up for the class.

As the date for the class approaches, a series of additional coordination emails are sent. Two weeks before the class starts, the HELM service sends a "Class logistics" email to the teacher, and corresponding email regarding the class to all registered students. A version of this email (with slight wording variations) is resent a week before the class, 3 days before, and a day before the class. On the day of the class, 10 minutes before the class, the HELM service automatically sends a "Join now" email to the teacher and students enrolled to ensure maximum class attendance.

Figure 10 further shows several coordination emails that are dynamically generated and sent to the students and teachers. This includes the automatically generated class recording links, and the reminders to join the class session before each class session. The class recordings are stored in a shareable Google Drive storage that is shared with the class students and the teacher. Finally, at the end of the class, students are asked to rate the class and the teacher on a few criteria, which is useful to personalize future class content.

Finally, the Machine Learning system is again used by the HELM service to get the top 3 class recommendations for each student that took this class. This is used to dynamically personalize a "Next class recommendations" email to students.

### 5.5. HELM Data collection service

The responses to class surveys, ratings, and other feedback received from students and teachers is stored by the HELM data collection service in the user_feedback_responses table of the HELM database, as shown in Figure 11. This is recorded when the user clicks on links that are automatically included in the HELM emails and webpages created by the HELM software services.

Additionally, as users open any of the emails generated by the HELM system, or click on the links embedded in these emails, the HELM data collection service logs these interactions in user_interactions log of the database. Similarly, clicking on any of the links in the HELM webpages are also logged in the user_interactions log of the database. All this data, and future class interactions tracking that the HELM system will store in its database form a very powerful data set that can be used by HELM's machine learning system to personalize the platform even further for peer-to-peer learning.

## 6. HELM Machine Learning system

HELM's increase of classes over the past months means that many students must search through all classes to find a class that is interesting to them. Some of them may lose interest and not sign up for any class. The goal of the third and final part of my project was to create an algorithm that will predict human behavior, i.e., what a user will choose next based on previous actions. The reason this has not been done previously in a peer-to-peer learning context is because predicting human behavior is one of the hardest things a Machine Learning Algorithm can do ("The Fundamental Attribution Error: Why Predicting Behavior is so Hard," 23 Aug. 2016).

As part of developing a machine learning algorithm to recommend the next class to a student, two engineering questions were posed. First, was it possible to create a Machine Learning model that will predict, with over 80% accuracy, what class a student will take next based on previous classes they have taken. Second, what was the best type of Machine Learning model to do such a task.

**6.1. Creating a ML model for student recommendation algorithm.** For this part of the project, a prototype model was created which then went through several iterations.

**Version 1:** The first version was to show all classes on the classes page. This worked originally, however as HELM got more classes, there were more and more classes that people had to choose from. This approach does not meet my performance criteria, i.e, > 80% accuracy of prediction of probability that a student will take a given class.

**Version 2:** Version 2 was showing the most popular classes to students. This model was based on the assumption that students will likely sign up for popular classes. However this means the most popular classes become more popular, and is, therefore, biased. This approach also did not meet the performance criteria and the performance accuracy was low (< 30%).

**Version 3:** Version 3 was to create a Machine Learning model where the input was previous classes a student has taken and output was the next 3 classes a student would take. This approach created one large model that included all the possible classes. This approach also did not meet the performance criteria as the accuracy was around ~60%.

**Version 4:** In version 4, an individual Machine Learning models for each class was created, where the input was the student data and the output was the probability that the student will take that class (see Figure 12). Specifically, the input Data was a list of classes that the student has previously taken and was formatted in a way that is friendly for the algorithm. For the model training, each student sign-up was used as a datapoint for each class. The probability was determined based on how their previous classes are similar to the class at hand (0-1). Further, similarity was based on how classes are tagged. The output data was the probability of taking a specific class. The fourth version is more accurate than the other 3 versions (~89% accuracy).

**6.2. Comparing different ML models for the best recommendation algorithm.** A second goal was to examine which Machine Learning (ML) model would work best to predict human behavior. Three possible ML models were identified based on research on similar topics. First, a Random Forest (Pedregosa et al., 2011) algorithm essentially uses a bunch of Decision Trees, which determine the outcome based on a bunch of decisions. The reason it uses several Decision Trees is because Decision Trees usually overfit data, meaning it draws a fine line around the training data, and does not accurately predict for any other inputs. Using several or "forest" Decision Trees means that the algorithm will not overfit. The second algorithm was a Logistic Regression (Brownlee, 31 Mar. 2016; *What is Logistic Regression*, 2022), which is one of the best algorithms to take in binary data (meaning just a series of 1s and 0s) and output a prediction. It trains based on a Logistic Function, indicating that it will train until the accuracy increase plateaus. The final algorithm examined was an MLP classifier (Pedregosa et al., 2011), which is a Neural Network. Neural Networks are similar to Random Forest and Logistic Regression, in that it has input nodes and an output node, however Neural Networks have hidden layers in between the Input and Output layers, which help fine-tune the prediction to get the most accurate answer. Neural Networks are the most accurate, however they do take a lot of time to train due to the Hidden Layers. Subsequently, one ML model was created for each class using each model type - Random Forest, Logistic Regression, and MLP Classifier.

The three models were compared with a control condition which recommended the most popular class (similar to version 2 of the model). The accuracy of each model for each type was obtained and these were averaged to find the accuracy for each model type. I also recorded how fast each model type was in terms of training the model. Figure 13 shows the difference of accuracy for each model as a whole, and also broken down by model class. Each model was around the same overall accuracy (~88%), while the MLP classifier was significantly slower than the RF and LR.

The results of the data analysis show that it is possible to create an algorithm to predict with over 80% accuracy. Further, all three models predict the probability that a student will take a class with over 80% accuracy. While all three ML models predict with high accuracy, the Logistic regression model is the fastest to train. So we pick the Logistic Regression model as the best ML model for predicting human behavior for peer-to-peer learning.

However, the results also show that the more popular a class is, the less accurate a model is. These results are acceptable for less popular classes, but that means that the more sign ups a particular class gets, the less times that class will be recommended correctly. This observation is pretty counter intuitive, since generally the more data a model has, the more accurate the model is. This downwards slope is not a good indication for such prediction models, so the final version of the model involved some addition data wrangling.

**6.3. Version 5**. **Final ML model with student and class tags.** Version 5 uses the same idea as the fourth version, however considers an interest system to better understand the connection between classes. Each class in HELM has been tagged with certain interests. The recommendation system then uses the tags each class has and detects to see how many similar tags there are. For example, the Microbiology and Geobiology classes have 2 tags in common, the Science and the Biology tags, so they are 50% similar, whereas the Microbiology and Chemistry classes have only 1 tag similar, making them 20% similar. The interests system is used in both the input and output. In terms of input, for each class model, the datapoints are all 900 students. The input data for each student takes the answer for the question "has the student taken this class?" and the answer for that would be 0 for if they haven't, and 1 for they have. However, the binary values only work if it does not include tags. If similarity tags are included, then the classes that weren't taken get the similarity percentage, comparing the similarity of that class to the interests associated with the students. That will go in place of the 0. For the output without tags, it is a simple 1 or 0 for if they will / if they have taken the class. With tags, it is 0/10, 2/10,5/10,7/10,8/10,10/10, mainly because the model is made for classification, so giving it multiple options should make it more accurate. Finally, the model outputs the probability that a given student will take a particular class (or the top N classes)(see Figure 14).

The new results are shown in the scatterplots in Figure 15, and compare the correlation between the number of user signups and probability of a class recommendation. The figures also compare the relationship with and without using tags in the inputs and outputs. Figure 15 shows that in the top two graphs, when the input doesn't use tags, accuracy is just as bad as version 4; the more popular a class gets, the less accurate the class model gets. However, looking at the averages, the figure shows that having inputs with tags help. To choose which of the bottom two graphs are the best, we can look at the averages and maximums of each graph and see that the right graph has better, and can be used for implementation. Thus, the lower right figure shows that having input with tags but outputs without helps.

Version 5 is the best version to use as it has the highest accuracy (higher than version 4 which was better than all other versions). The best model type to use is the Logistic Regression as it has a high accuracy, with a very low training time, indicating it will work the best when implemented in the HELM Learning platform (see Figure 16).

**7. Conclusion**

This paper describes the development and validation of world's first peer-to-peer online education platform, HELM (Helping Everyone Learn More), where any student can teach other students subjects or topics that they have mastered. HELM was created in 2020 when the pandemic started and structured education was disrupted in most schools. As of September 2022, the platform hosts 80 classes with over 4000 student sign ups across 4 continents. The HELM platform is built using a LAMP (Linux, Apache, MySql, Python) software stack and hosted on AWS (Amazon Web Services), so that it is available to everyone online. The HELM platform provides a process for "student teachers" to create classes they would like to teach, which is made available online if the teacher and class meets appropriate quality criteria. The platform automatically recommends appropriate classes to the right students, sets up the class logistics, sends automatic reminders to enrolled students, and ensures the classes are completed by students.

The data collected about the classes, students and teachers is used by HELM's Machine Learning (ML) system to make class recommendations. Several ML algorithms were trained and tested: Random Forest,

Logistic Regression, and MLP neural network classifier. All three models had a high prediction accuracy for the next class recommended to a student, while the Logistic Regression model has the least training time. As we make HELM available to more people, we will study additional ML models that predict the behavior of students and teachers to help even more students teach and learn more. Finally, what can we learn about predicting human behavior from this? As in the analysis for Version 5, using interest tags help. This is because it gives the computer a way to tell that two classes are very close and should be associated with one another. The other thing important outcome - as seen in the experiment for Version 4 - is that the best models to use are the Logistic Regression and Random Forest, as they are fast and very accurate for doing the task. Version 5 is implemented in to the HELM system currently, such that it can accurately recommend to the next student (e.g., Alice) what top three classes she/he should take.

Figure 1: Helping Everyone Learn More (HELM) website's main webpage

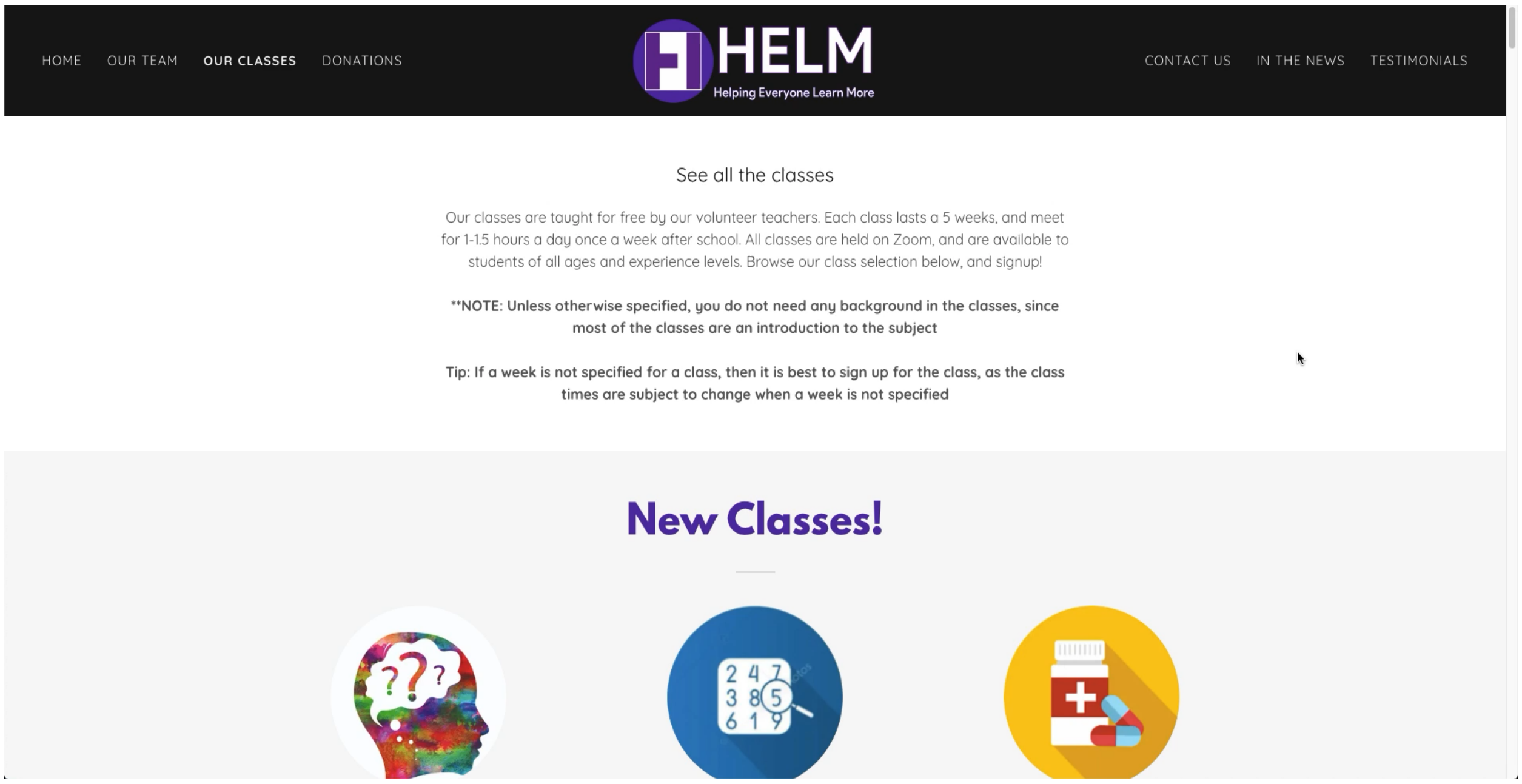

Figure 2: HELM classes webpage showing a sample of HELM classes

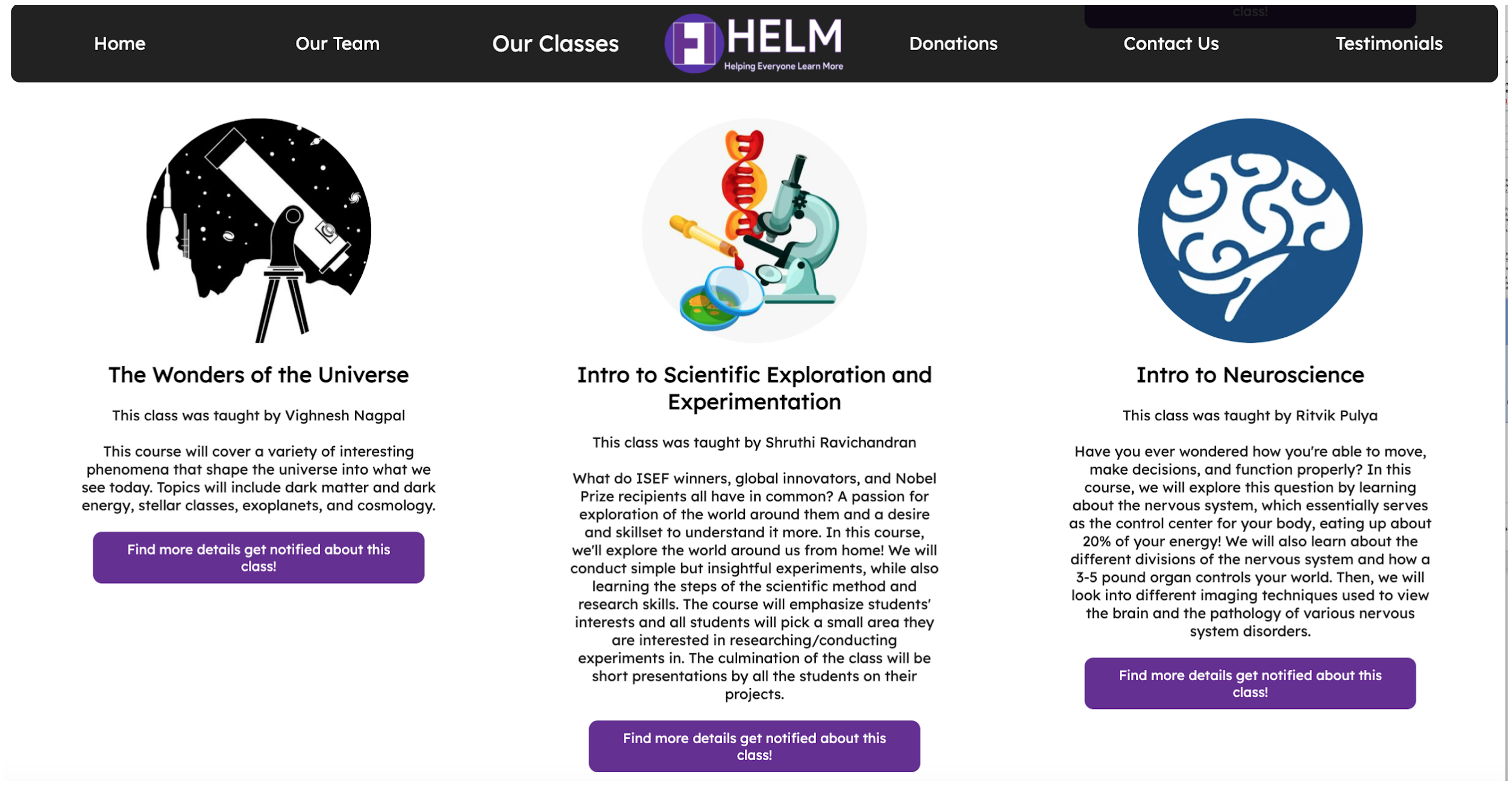

Figure 3: HELM teachers webpage showing some HELM teachers

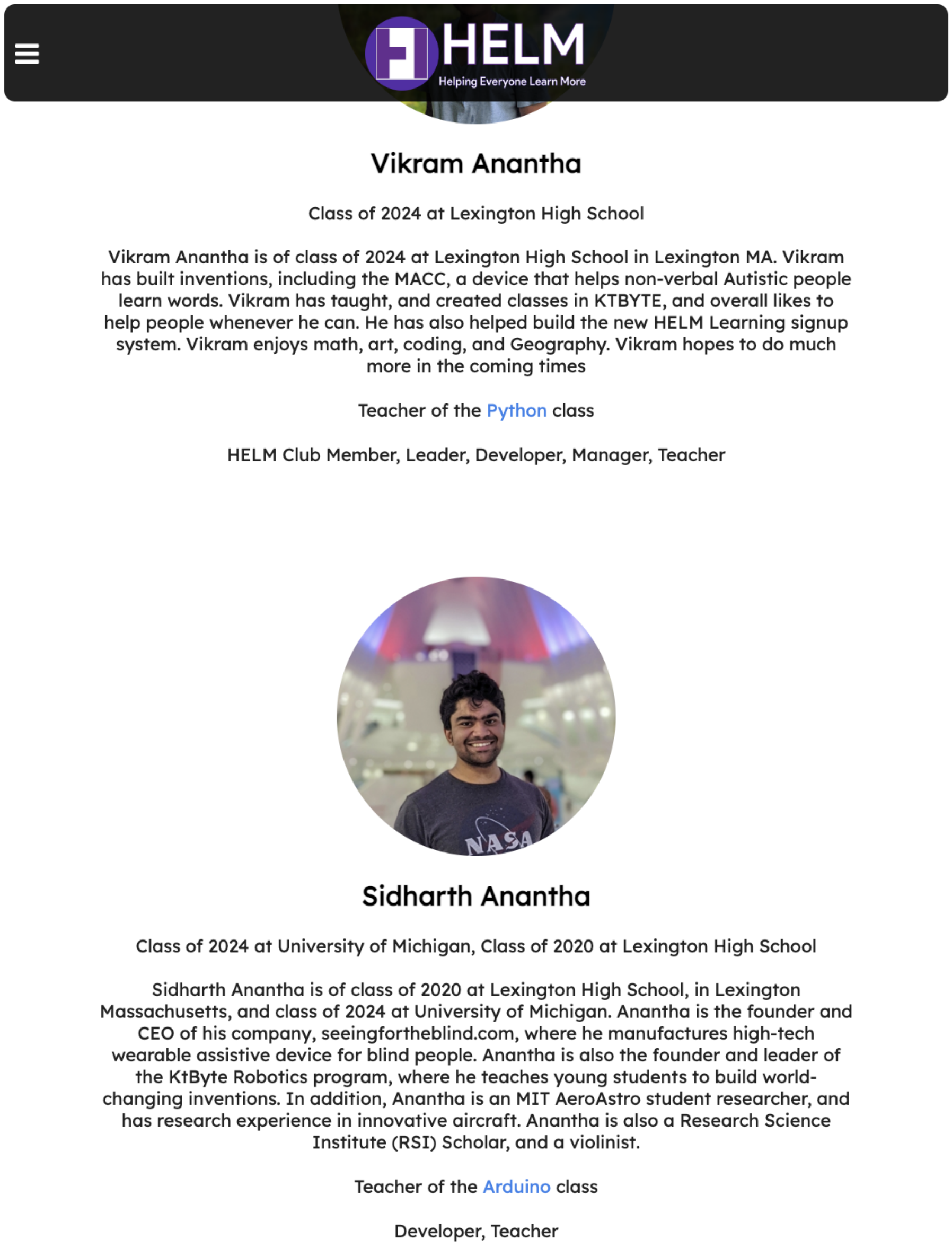

Figure 4: HELM Software Platform components including the HELM frontend system, HELM backend systems, and the HELM data store.

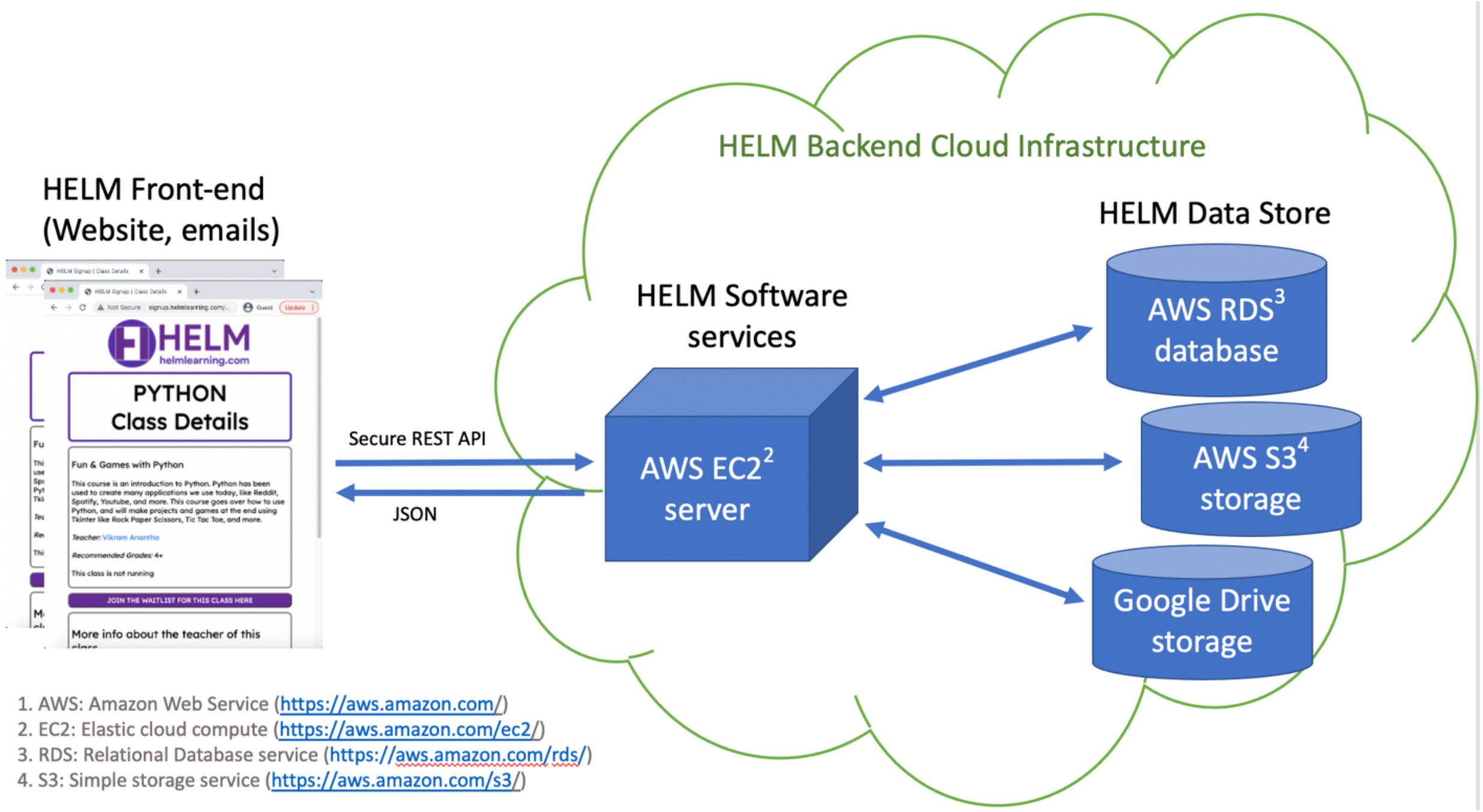

Figure 5: HELM Software Services (backend) and its interaction with different types of HELM users.

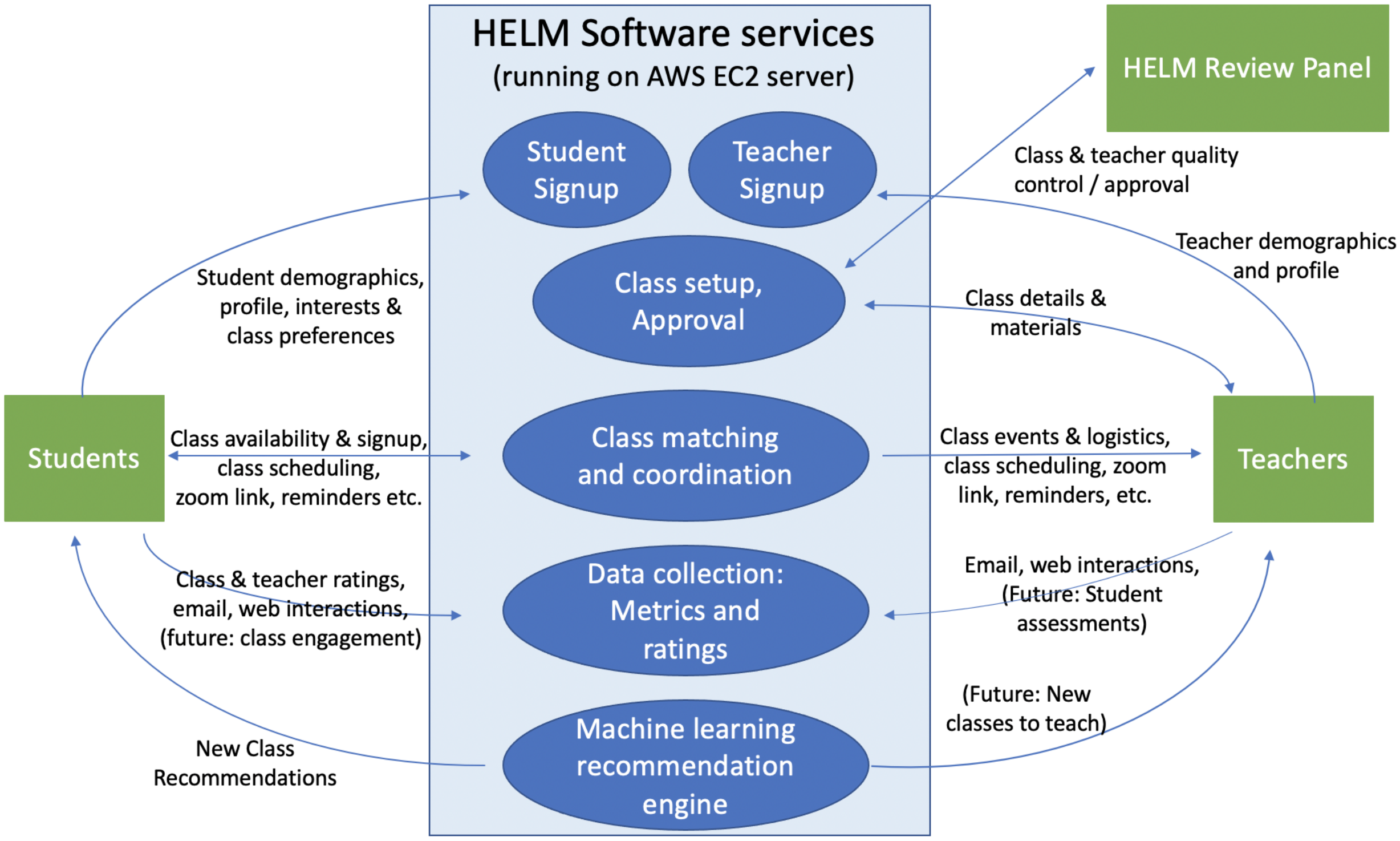

Figure 6: HELM *Student signup software service* – algorithm and the data flow example when a new student registers with HELM.

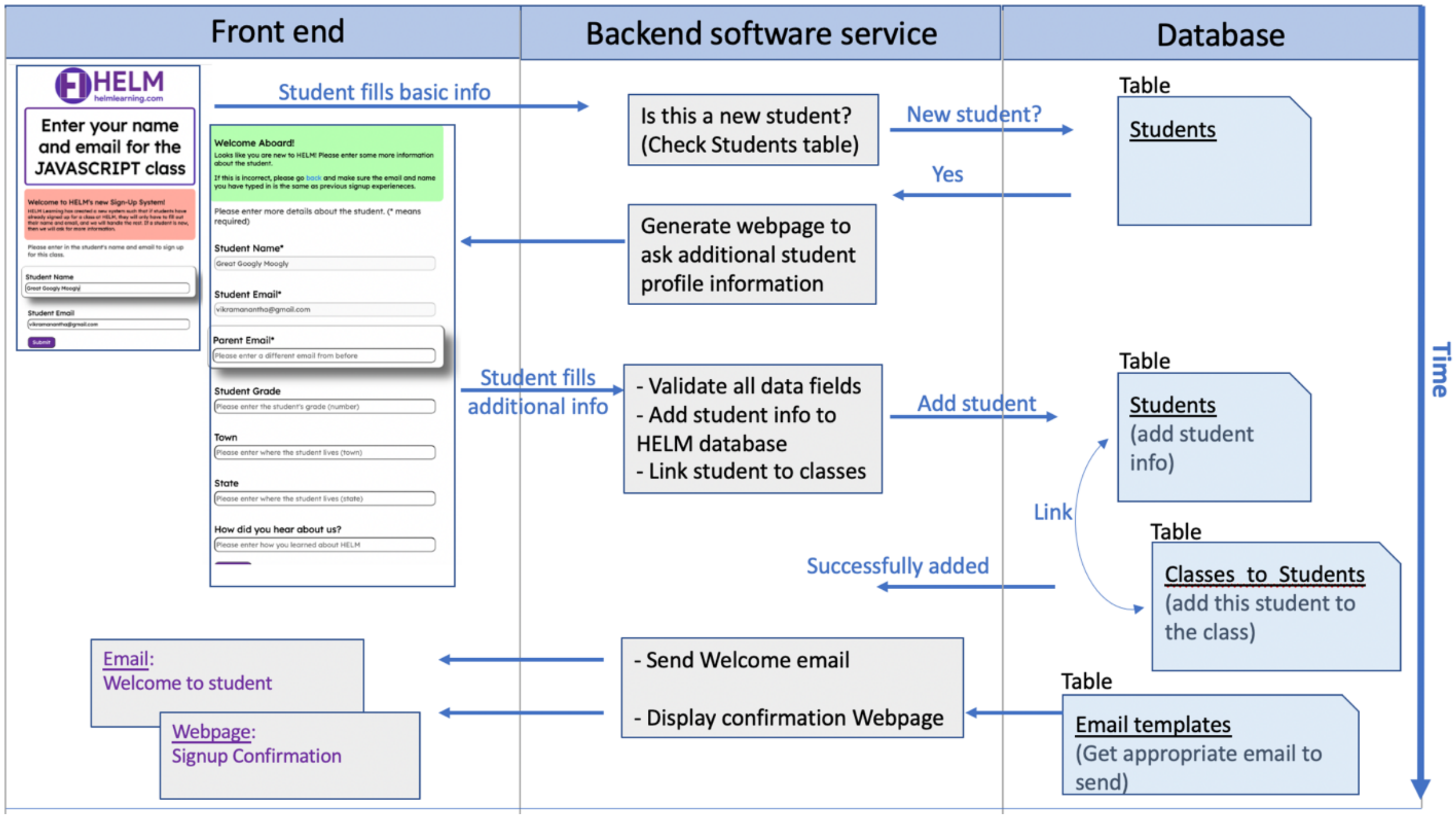

Figure 7: HELM *Teacher signup service* – algorithm and the data flow example when a new teacher wants to teach a new class.

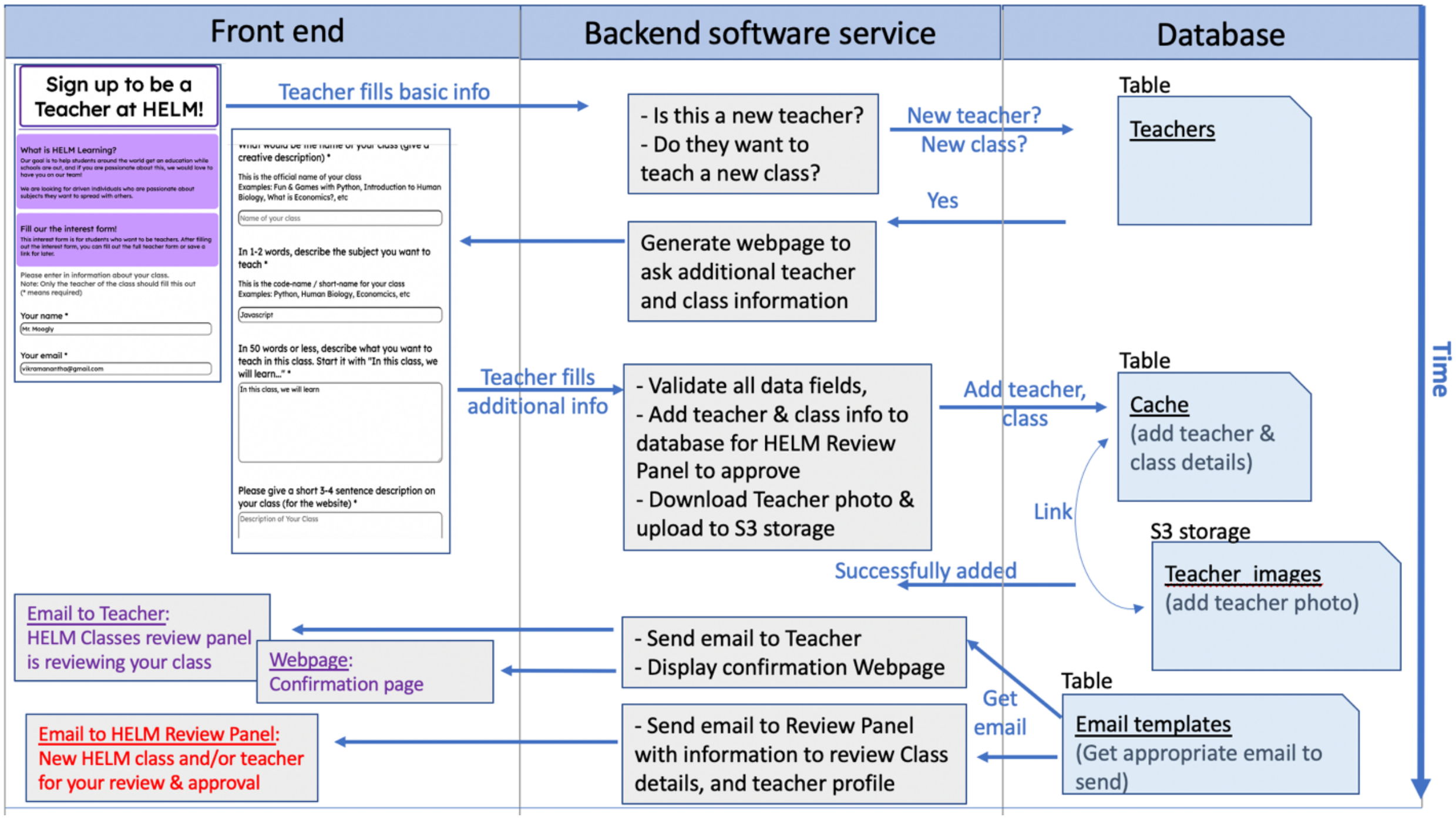

Figure 8: HELM *Class setup & approval service* – algorithm and data flow example for the approval of a new teacher and class.

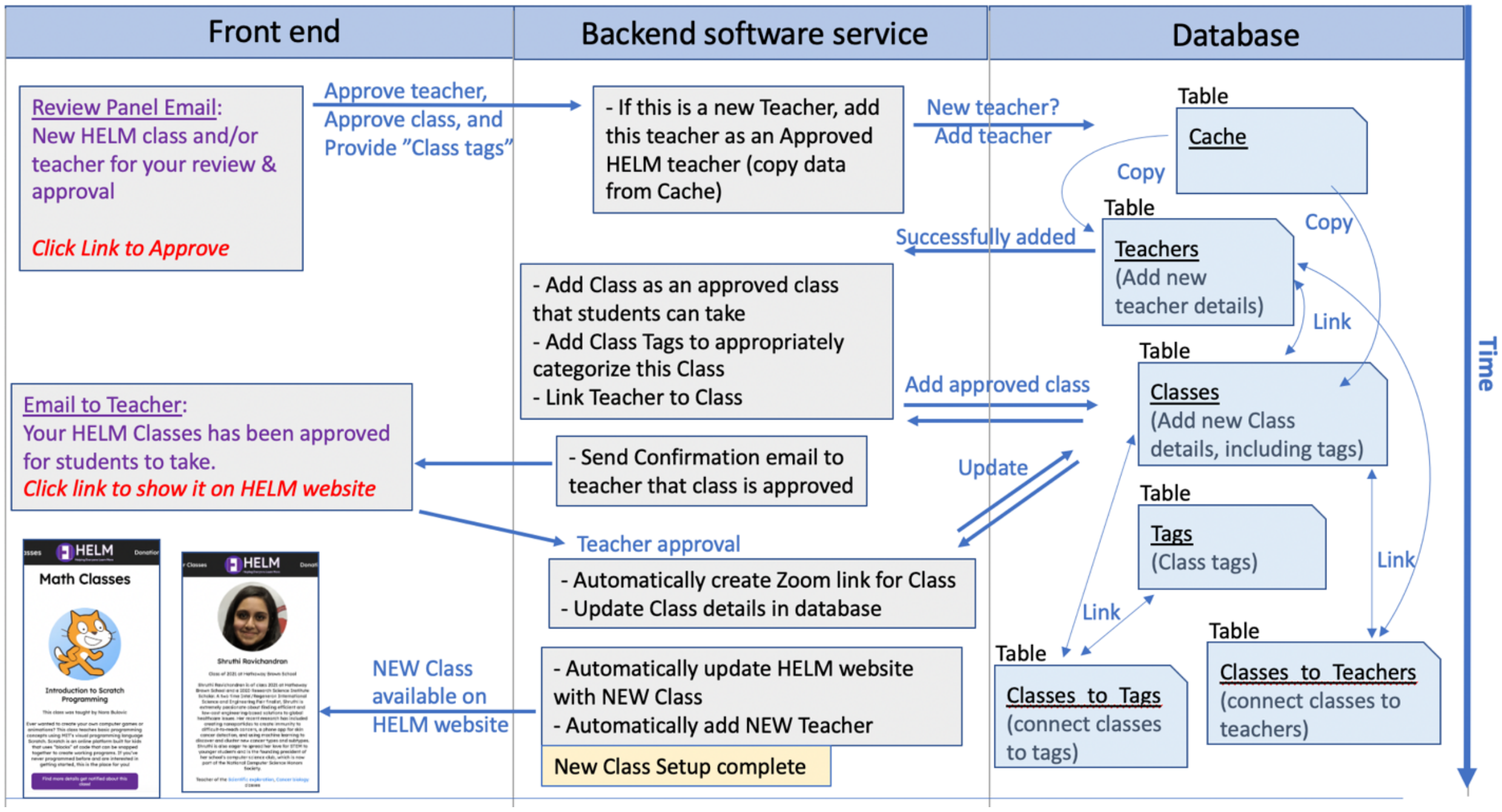

Figure 9: HELM *class matching and coordination service* (part 1 of 2) – algorithm and data flow example of the matching and coordination for a new class that has just been setup.

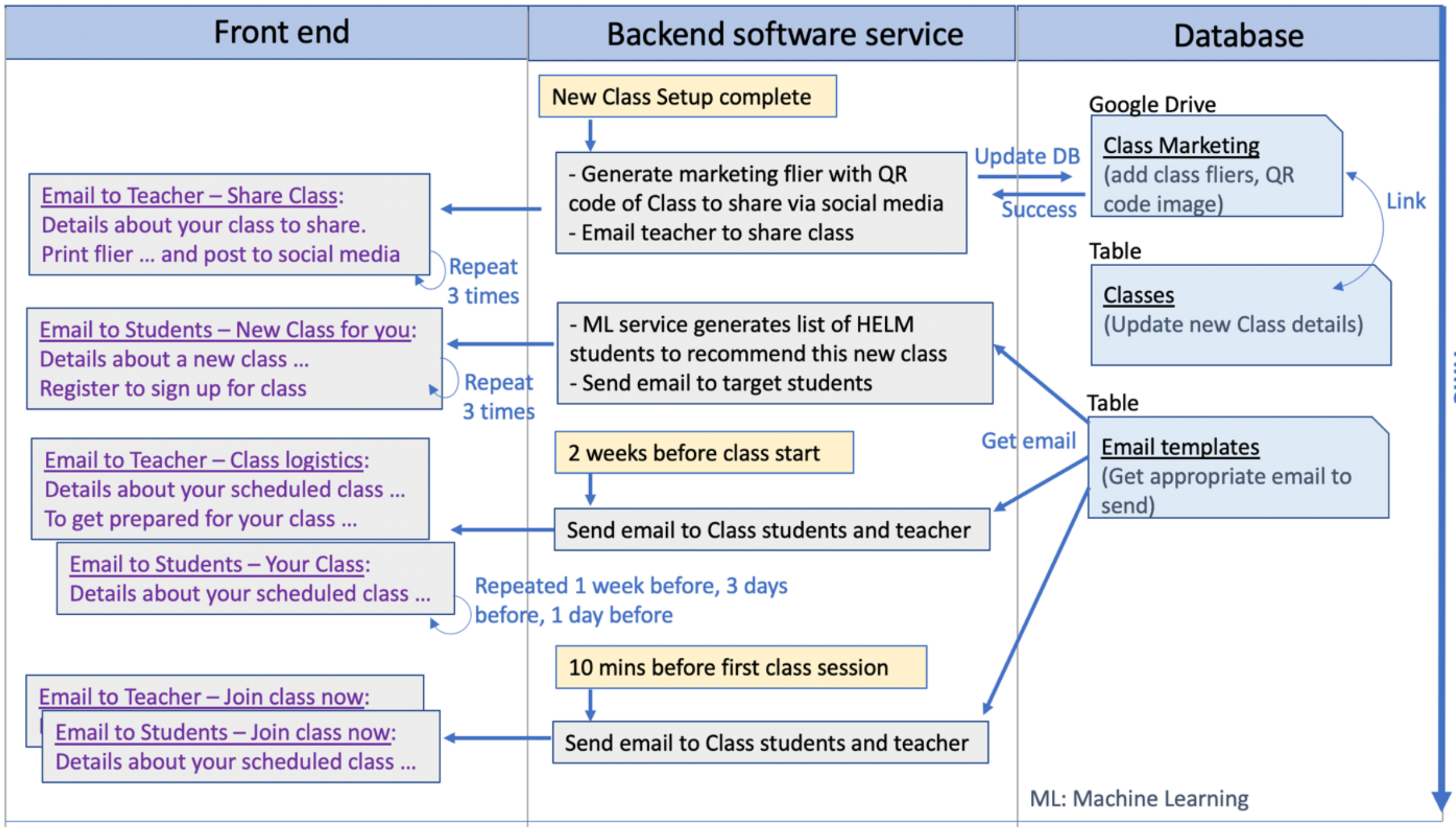

Figure 10: HELM *class matching and coordination service* (part 2 of 2) – algorithm and data flow example of the matching and coordination for a new class that has just been setup.

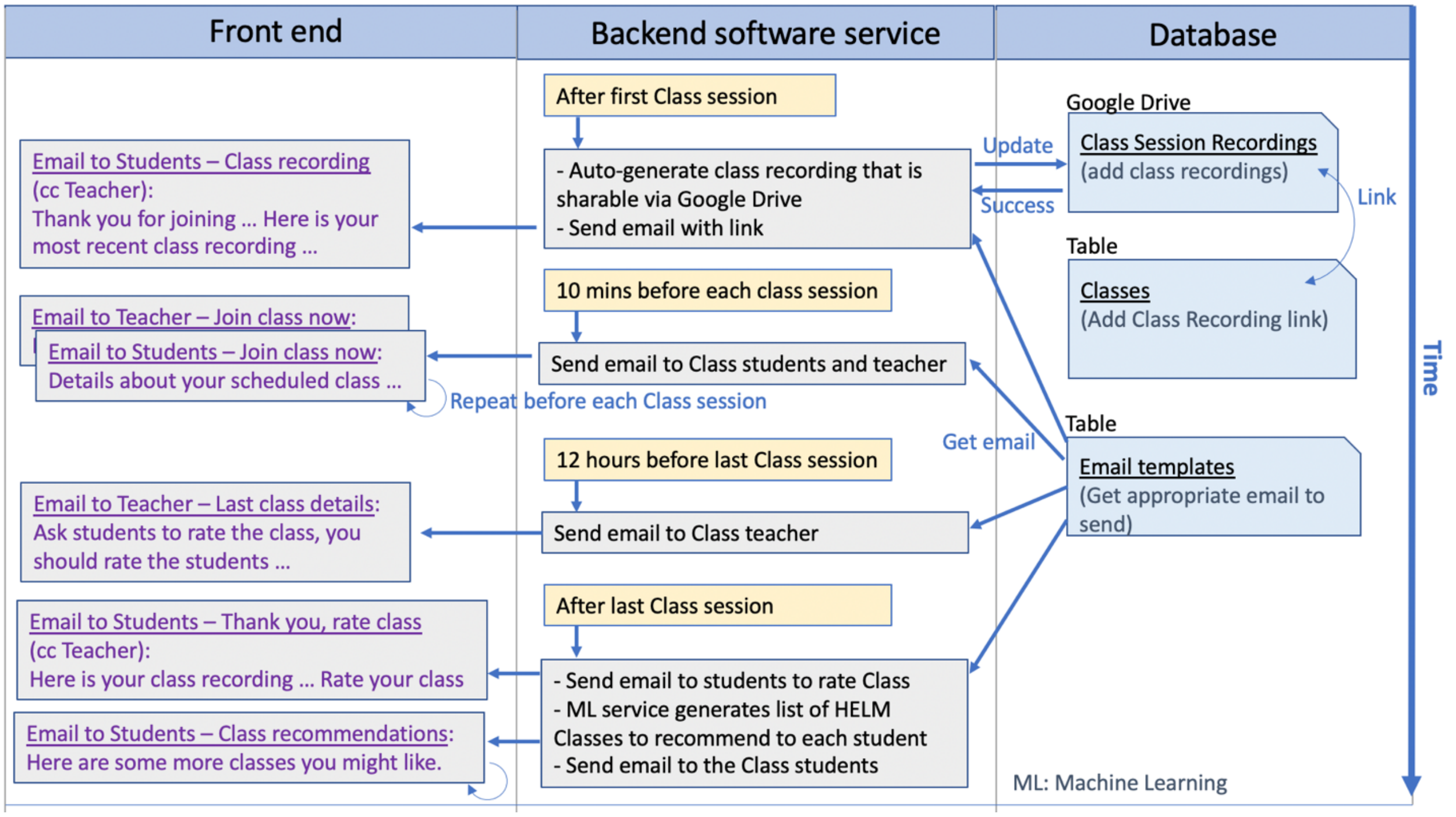

Figure 11: HELM *data collection service* – example of the types of data collected from student and teacher interactions that is stored in the HELM database.

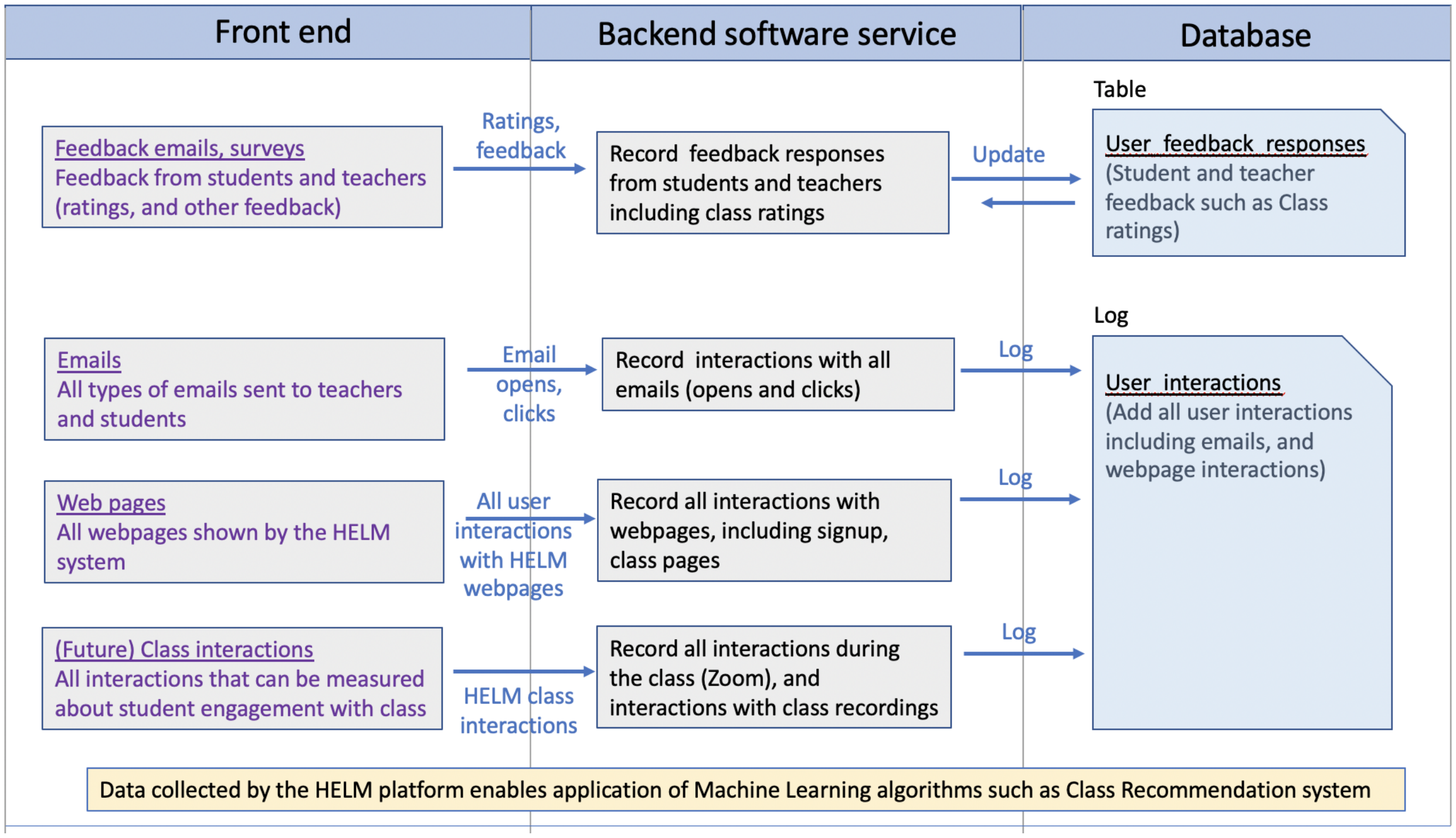

Figure 12: Individual Machine Learning models for each class was created for version 4, where the input was the student data and the output was the probability that the student will take that class

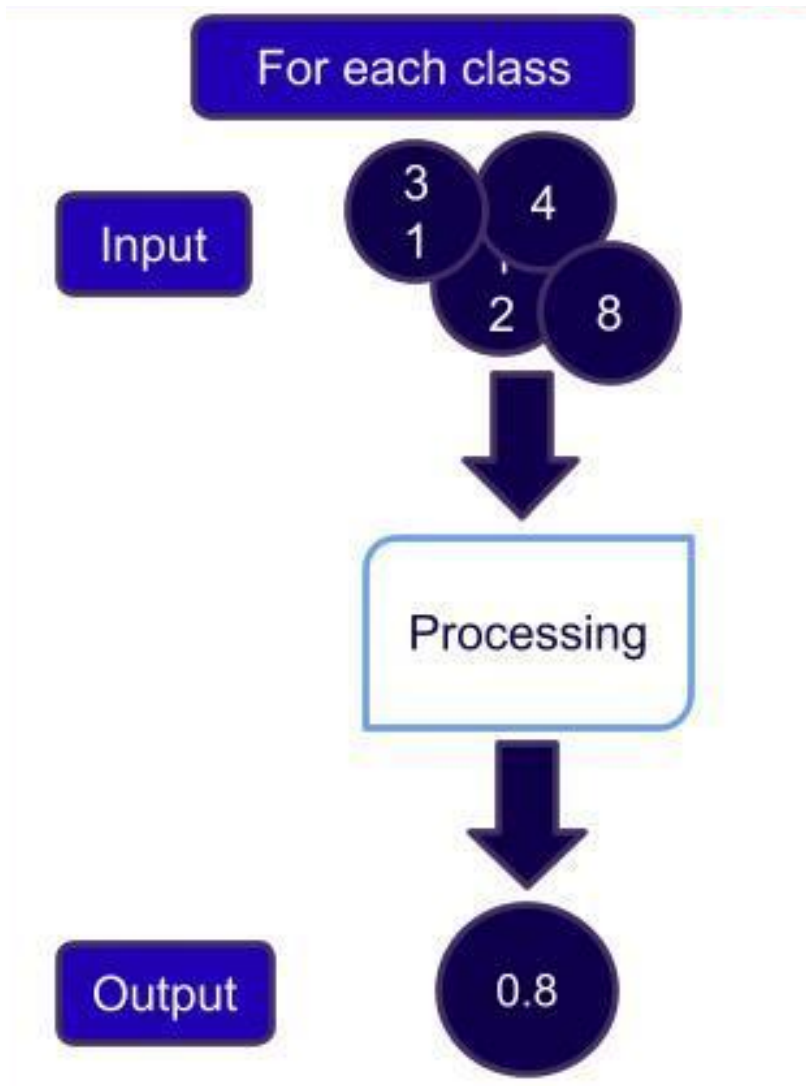

Figure 13: Average accuracy for each of the ML models relative to a control model (version 2) showing the MLP classifier is the most accurate (a) but also most time consuming (b).

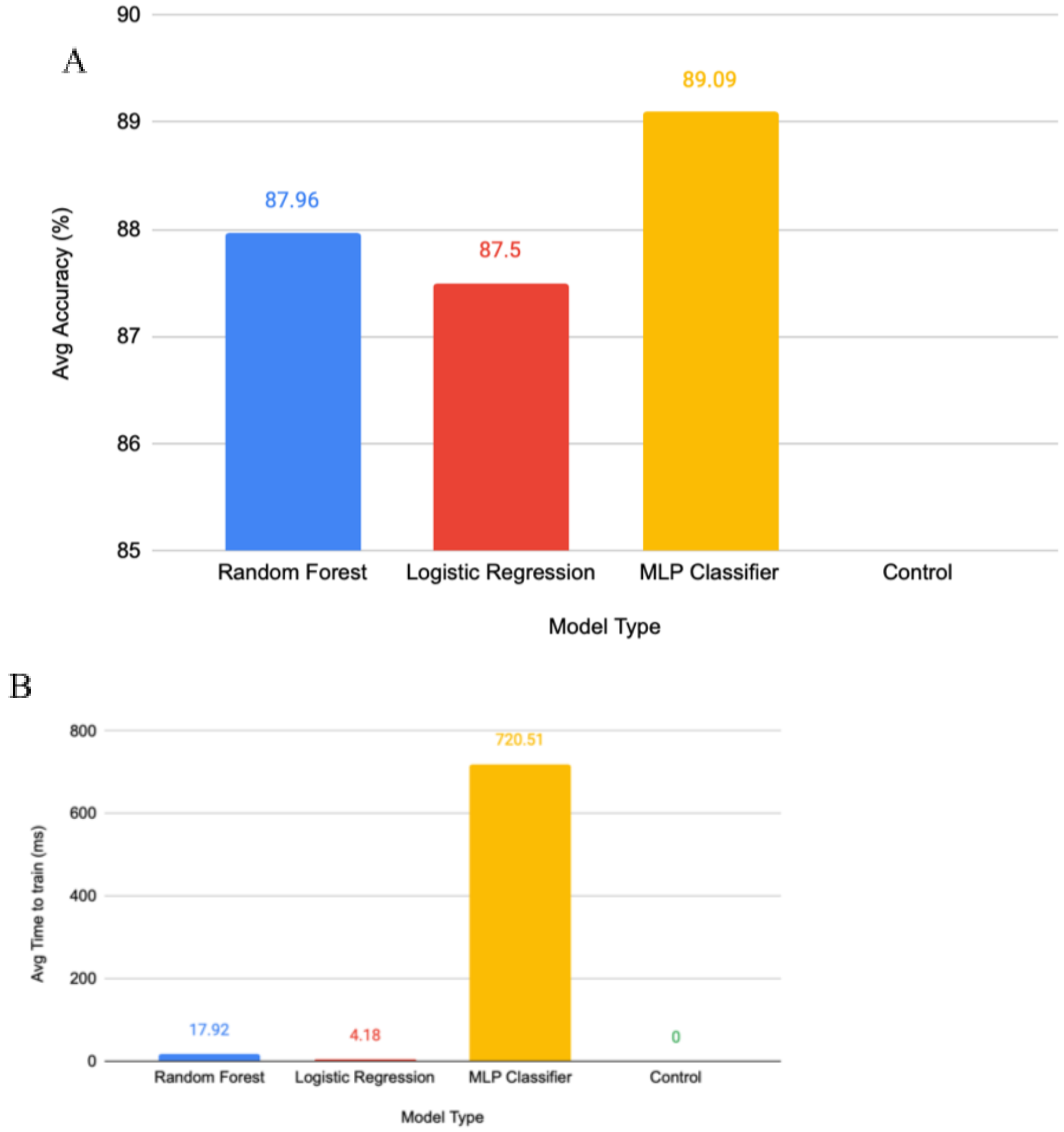

Figure 14: Version 5 is a modified version of Version 4, as it considers an interest based system to better understand the connection between classes.

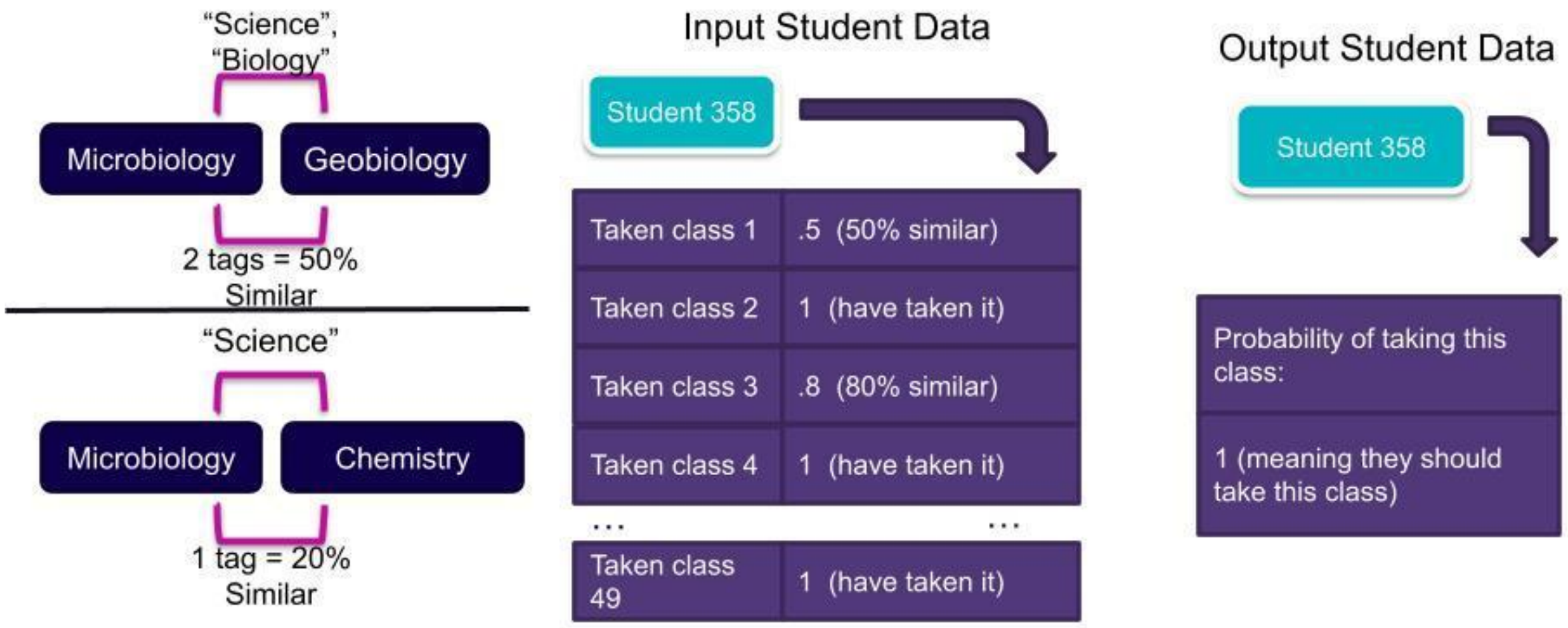

Figure 15. This figure shows the correlation between number of unique signups and accuracy for model prediction. The lower right figure shows that having input with tags but without outputs helps the overall model prediction performance.

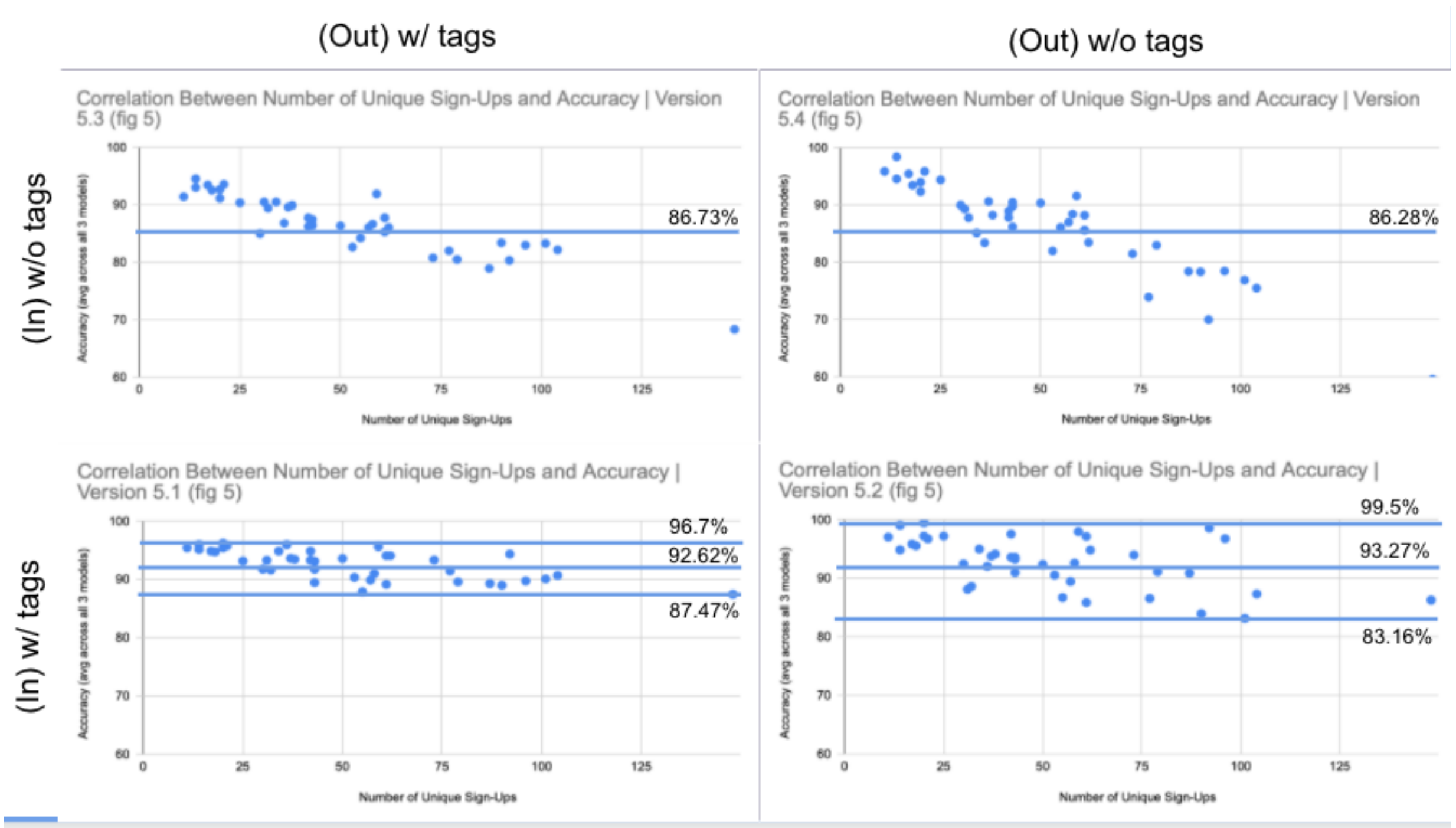

Figure 16. This figure shows that version 5 ML models have higher accuracy than version 4 models (A). This figure also shows that the best model type to use is the Logistic Regression as it has a high accuracy, with a very low training time (B), meaning it will work the best when implemented in the HELM Learning platform.

A

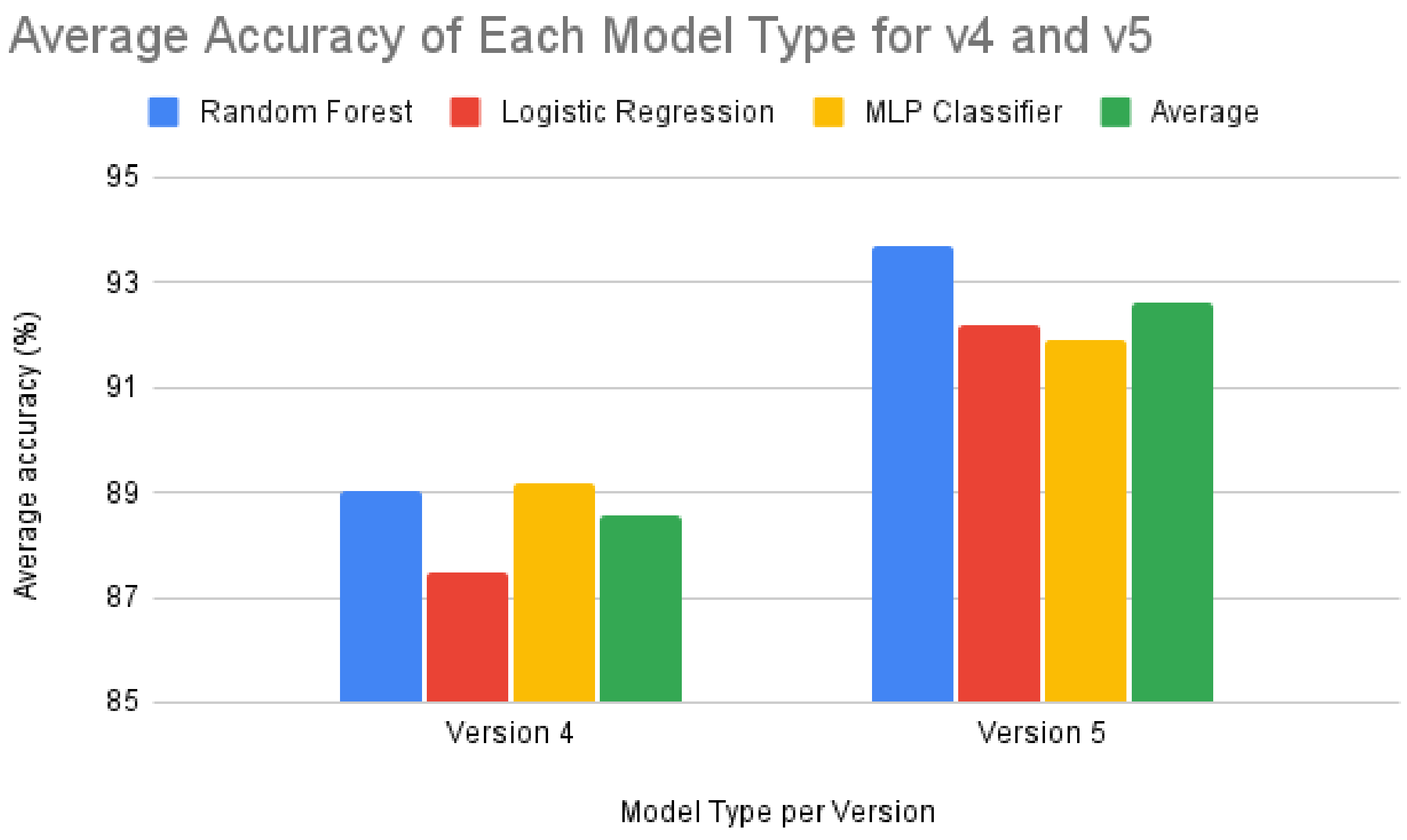

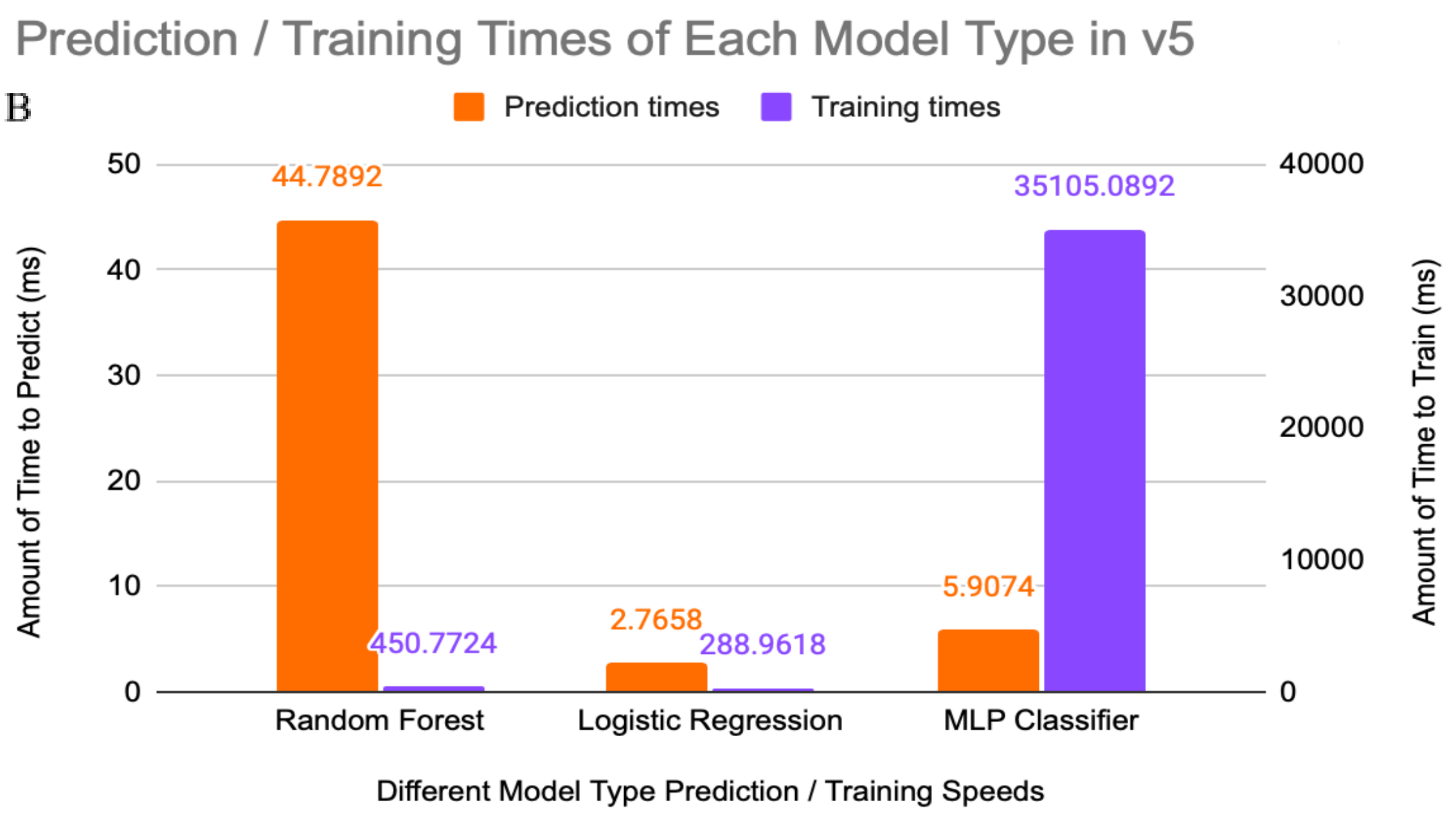
Prediction / Training Times of Each Model Type in v5
B
Prediction times
Training times
Amount of Time to Predict (ms)
Amount of Time to Train (ms)
50
40
30
20
10
0
40000
30000
20000
10000
0
44.7892
450.7724
2.7658
288.9618
5.9074
35105.0892
Random Forest
Logistic Regression
MLP Classifier
Different Model Type Prediction / Training Speeds